\documentclass[11pt]{article}
\usepackage{amsmath,amssymb,color,cite}
\usepackage{graphicx}
\usepackage{amsfonts}
\usepackage{enumerate}
\usepackage{bbm}
\usepackage{multirow}
\usepackage{nicefrac}
\usepackage[all]{xy}
\usepackage{graphicx}
\usepackage{booktabs}

\usepackage{amsfonts}
\usepackage{comment}
\usepackage{slashed}

\def\b{{\beta}}

\def\oneone{\rlap 1\mkern4mu{\rm l}}

\newcommand{\w}[1]{\\[0.#1cm]}
\newcommand*{\itemequation}[3][]{%
  \item
  \begingroup
    \refstepcounter{equation}%
    \ifx\\#1\\%
    \else  
      \label{#1}%
    \fi
    \sbox0{#2}%
    \sbox2{$\displaystyle#3\m@th$}%
    \sbox4{\@eqnnum}%
    \dimen@=.5\dimexpr\linewidth-\wd2\relax
    \ifcase
        \ifdim\wd0>\dimen@
          \z@
        \else
          \ifdim\wd4>\dimen@
            \z@
          \else 
            \@ne
          \fi 
        \fi
      \@latex@warning{Equation is too large}%
    \fi
    \noindent   
    \rlap{\copy0}%
    \rlap{\hbox to \linewidth{\hfill\copy2\hfill}}%
    \hbox to \linewidth{\hfill\copy4}%
    \hspace{0pt}
  \endgroup
  \ignorespaces 
}

\makeatletter
\@addtoreset{equation}{section}
\makeatother
\renewcommand{\theequation}{\thesection.\arabic{equation}}

\newcommand{\be}{\begin{equation}}
\newcommand{\ee}{\end{equation}}
\newcommand{\bea}{\setlength\arraycolsep{2pt} \begin{eqnarray}}
\newcommand{\eea}{\end{eqnarray}}
\newcommand{\nn}{\nonumber}

\def\bea{\begin{eqnarray}}
\def\eea{\end{eqnarray}}
\newcommand{\ft}[2]{{\textstyle\frac{#1}{#2}}}
\def\a{\alpha}
\def\b{\beta}

\def\G{\Gamma}
\def\d{\delta}

\def\L{\Lambda}

\def\o{\omega}
\def\O{\Omega}

\def\tr{\rm tr}

\def\vp{{\varphi}}

\def\eq#1{(\ref{#1})}

\begin{document}

\begin{flushright}
\hfill{MI-TH-202}\\

\today

\end{flushright}

\vspace{25pt}

\begin{center}
{\Large         {\bf On the consistency of a class of $R$-symmetry \\gauged
$6D\ {\cal N}=(1,0)$ supergravities}     }

\vspace{0.3in}

{\large
Yi Pang$^1$ and Ergin Sezgin$^2$
}

\vspace{0.3in}

{$^1$ \it Mathematical Institute, University of Oxford, 
Woodstock Road, Oxford OX2 6GG, U.K. }

\vspace{10pt}

$^2${\it George and Cynthia Woods Mitchell  Institute
for Fundamental Physics and Astronomy,\\
Texas A\&M University, College Station, TX 77843, USA}

\vspace{40pt}

\begin{abstract}

$R$-symmetry gauged $6D\, (1,0)$ supergravities free from all local anomalies, with gauge groups $G\times G_R$ where $G_R$ is the R-symmetry group and $G$ is semisimple with rank greater than one, and which have no hypermultiplet singlets, are extremely rare. There are three such models known in which the gauge symmetry group is $G_1\times G_2 \times U(1)_R$ where the first two factors are $ \left(E_6/{\mathbb{Z}_3}\right) \times E_7$, $ G_2 \times E_7 $ and $F_4 \times Sp(9)$. These are models with single tensor multiplet, and hyperfermions in the $(1,912)$, $(14,56)$ and $(52,18)$ dimensional representations of $G_1\times G_2$, respectively. So far it is not known if these models follow from string theory.  We highlight key properties of these theories, and examine constraints which may arise from the consistency of the quantization of anomaly coefficients formulated in their strongest form by Monnier and Moore. Assuming that the gauged models accommodate dyonic string excitations, we find that these constraints are satisfied only by the model with the $F_4 \times Sp(9)\times U(1)_R$ symmetry. We also discuss aspects of dyonic strings and potential caveats they may pose in applying the stated consistency conditions to the $R$-symmetry gauged models.

\end{abstract}

\end{center}

\vspace{15pt}

\thispagestyle{empty}

\pagebreak
\voffset=-40pt



\section{Introduction}

This paper is dedicated to Michael Duff on the occasion of his 70th birthday. The work described here is on a subject in which Michael has made magnificent contributions. Let us also remember that his advocacy of supermembranes and eleven dimensions, prior to their wide acceptance, is in the annals of physics. He has related amusing anecdotes about the era prior to this acceptance. One of us (ES) has this one to add: In an Aspen Conference in 1987, in a conversation on supermembranes and 11D, a very distinguished colleague said that ``in Cambridge meeting there were 5 hours of talk on the subject, this must be a Thacherian plot to destroy the British physics"! Asked about 11D supergravity,  he replied , ``it is a curiosity"!

In the spirit of exploring some other ``curiosities", here we aim at drawing attention to a class of supergravities in six dimensions that are free from all local anomalies in a rather remarkable fashion. To begin with, let us recall that the requirement of anomaly freedom has considerably constraining consequences for supergravity theories. The gauge groups and matter content are restricted, and on-shell supersymmetry, in the presence of the Green-Schwarz anomaly counterterm, {\it requires} the introduction of an infinite tower of higher derivative couplings.  
While this is not expected to fix uniquely an effective theory of quantum supergravity \cite{Bergshoeff:1989de,deRoo:1992zp,deRoo:1992sm,Tseytlin:1995bi},
it may nonetheless provide a framework for a $``\alpha'$-deformation" program in which extra consistency requirements, including those arising from global anomalies, may restrict further the effective theory. If consistent such theories exist, in principle they may offer viable spots in a region outside the string lamp post in a search for UV completeness. In fact, even in $10D$ it would be instructive to determine if and how the $\alpha'$ deformation approach runs into problems unless it is uniquely determined by string theory. 
If all roads lead to string  theory that too would be a valuable outcome in this program, providing  more evidence for what is referred to as the ``string lamp post principle" \cite{Kim:2019ths,Palti:2019pca}.

In this note, as mentioned above, we draw attention to a class of 6D supergravities which are remarkably anomaly free, and yet it is not known if they can be embedded into string theory. These are $U(1)_R$ gauged supergravities with specific gauge groups, and from the string theory perspective unusual hypermatter content. The qualification `remarkable' is due to the fact that $R$-symmetry gauged $6D\, (1,0)$ supergravities free from all local anomalies, and with gauge groups $G\times G_R$ where $G_R$ is the R-symmetry group and $G$ is semisimple with rank greater than one, and without any hypermultiplet singlets, are extremely rare; see for example \cite{Avramis:2005hc}. 
By contrast, there is a huge number of  anomaly free  ungauged $6D,\ (1,0)$ supergravities\footnote{There is a crucial difference between gauging of $R$-symmetry compared to gauging of non R-symmetries. Here, we shall reserve the word ``gauged" to mean ``$R$-symmetry gauged".}
one can construct directly, and  a very large class that can be embedded, indeed directly be obtained from, string theory; see for example \cite{Green:1984bx}. One can also find several $R$-symmetry gauged models in which there are several hypermultiplet singlets, and gauge group G that involves a number of $U(1)$ factors \cite{Avramis:2005hc,Suzuki:2005vu}. 
 
If we insist on semisimple gauge groups and exclude hyperfermion singlets, then there are only three anomaly free gauged $6D\, (1,0)$ supergravities known so far. They have the  gauge symmetry $G_1\times G_2 \times U(1)_R$ where the last factor is the gauged $R$-symmetry group, and the first two factors are $ \left(E_6/{\mathbb{Z}_3}\right) \times E_7$ \cite{RandjbarDaemi:1985wc}, $ G_2 \times E_7 $ \cite{Avramis:2005qt}  and $F_4 \times Sp(9)$ \cite{Avramis:2005hc}. These are models with a single tensor multiplet, and hyperfermions in the $(1,912)$, $(14,56)$ and $(52,18)$ dimensional representations of $G_1\times G_2$, respectively. The embedding of these theories in string theory is not known \footnote{If one considers strictly the $U(1)_R$ gauged theory with $n_T=1$ and no other gauge sector and hypermatter \cite{Salam:1984cj},  it has been shown \cite{Cvetic:2003xr} to follow from pure Type I supergravity in 10D, on a  non-compact hyperboloid ${\cal H}_{2,2}$ times $S^1$ and a consistent chiral truncation. However, the inclusion and origin the Yang-Mills hypermatter remains an open problem.}. 
In particular, the $\left(E_6/{\mathbb{Z}_3} \right)\times E_7 \times U(1)_R$ model contains a representation of $E_7$  beyond the fundamental and adjoint what normally one encounters in string theory constructions. Neither $R$-symmetry gauging, nor such representations seem to arise in string/F theory constructions \footnote{It has been suggested in \cite{Distler:2007av} that higher Kac-Moody level string worldsheet  algebras may lead to such representations but this matter is far from being settled.}. 
Another landmark of these models is that they come with a positive definite potential proportional to the exponential function of the dilaton, even in the absence of the hypermultiplets. This has significant consequences. For example, these models do not admit maximally symmetric $6D$ spacetime vacua, but rather Minkowski$_4 \times S^2$ with a monopole configuration on $S^2$ \cite{Salam:1984cj}. In its simplest form, such gauged supergravities seem to have attractive features for cosmology (see, for example,  \cite{Maeda:1985es,Halliwell:1986bs,Gibbons:1986xp,Gibbons:1987ps,Anchordoqui:2019amx}).  The presence of the potential also has interesting consequences for the important question of whether dyonic string excitations  are supported by the theory. While some encouraging results have been obtained in that direction \cite{Gueven:2003uw,RandjbarDaemi:2004qr,Jong:2006za}, much remains to be investigated.

The Green-Schwarz mechanism ensures  the cancellation of local anomalies.  Demanding the absence of possible global anomalies, on the other hand, can impose additional constraints. Such anomalies can arise from different aspects of the data furnished by the local anomaly free theory, and they can be rather difficult to compute. The best understood global anomaly has to do with large gauge transformations not connected to the identity. The models in question are free from these anomalies. Other global considerations, motivated in part by lessons learned from the F-theory construction of anomaly free 6D theories \cite{Kumar:2010ru},  have led to additional constraints. 

To begin with, Seiberg and Taylor \cite{Seiberg:2011dr}, employed the properties of the dyonic charge lattice and the Dirac quantization conditions they must satisfy, to deduce the consequences for the coefficients of the anomaly polynomial. They observed that these coefficients form a sublattice of dyonic string charge lattice, and that the consistency requires that this can be extended to a unimodular (self-dual) lattice.  
A stronger condition was put forward relatively recently by Monnier, Moore and Park \cite{Monnier:2017oqd} who assumed that a consistent supergravity theory may be put on an arbitrary spin manifold and that any smooth gauge field configurations should be allowed in the supergravity ``path integral'', referring to this assumption as the ``generalized completeness hypothesis". They find a constraint which states that the anomaly coefficients for the gauge group $G$ should be an element of $2 H^4(BG; \mathbb{Z})\otimes \Lambda_S$ where $BG$ is the classifying space of the gauge group $G$, and $\Lambda_S$ is the unimodular string charge lattice.  

Monnier and Moore \cite{Monnier:2018nfs,Monnier:2018cfa} have further argued that these constraints are necessary but not sufficient for the cancellation of all anomalies, local and global, and proposed the necessary and sufficient conditions. They do so by requiring that the Green-Schwarz anomaly counterterm is globally well  defined, and show that this leads to the requirement that for any given GS counterterm, there must exist a certain 7D spin topological field theory that is trivial.  
These authors arrive at a proposition in \cite{Monnier:2018nfs} which states the conditions that need to be satisfied for the 6D theory to be free from all anomalies, in the case of connected gauge groups. Assuming that the theories under consideration here support dyonic string excitations, we will take this proposition as a basis for testing the consistency of these theories. We will see that the model based on $\left(E_6/{\mathbb Z}_3\right)\times E_7\times U(1)_R$ satisfies the weaker set of constraints mentioned above, but not all the conditions of the stronger criterion of Monnier and Moore, and that the model based on $F_4\times Sp(9)\times U(1)_R$ remarkably satisfies them all.

The paper is organized as follows. In the next section we shall describe the anomaly freedom aspects of the class of models being considered here. In Section 3, we recall the structure of the bosonic field equations, discuss the issue of higher derivative corrections, and survey the key properties of the few known dyonic string solutions. In section 4, we summarize the Monnier-Park constraints, and in section 6 we study these constraints for the models at hand. In the conclusions, we discuss the possible caveats in the interpretation of our results, and the appendix contain useful formula for the anomalies of the models under consideration,and in particular, we provide more details for the based on $F_4\times Sp(9)\times U(1)_R$ gauge group.

\section{$U(1)_R$ gauged anomaly free models}

We shall focus on $U(1)_R$ gauged $6D\, {\cal N}=(1,0)$ supergravities coupled to one tensor multiplet, vector multiples associated with group $G=G_1\times G_2\times U(1)_R$, and (half)hypermultiplets transforming in $(R_1,R_2)_0$ representation of $G_1\times G_2$, with the subscript denoting the $U(1)_R$ charge of hyperfermions.  The gravitino, dilatino and gauginos have unit $U(1)_R$ charge. The three models we shall consider have the gauge groups and hyperfermion contents  \cite{RandjbarDaemi:1985wc,Avramis:2005qt,Avramis:2005hc}
\begin{align}
& (A)  && \left( E_6/{\mathbb Z}_3\right) \times E_7 \times U(1)_R && (1,912)_0
\\
& (B)  && G_2 \times E_7 \times U(1)_R && (14,56)_0
\\
& (C)  && F_4 \times Sp(9) \times U(1)_R && (52,18)_0
\end{align}
The perturbative gravitational, gauge and mixed anomalies are encoded in an 8-form anomaly polynomial $I_8$, and they are cancelled by Green-Schwarz mechanism that exploits its factorization as
\be
 I_8 = \frac12\Omega_{\a\b} Y^\a Y^\b\ ,\qquad \O_{\a\b}=\begin{pmatrix} 0 & 1\\ 1 & 0 \end{pmatrix}\ ,
\ee
where
\be
Y^\a = \frac{1}{16\pi^2} \left( \frac12 a^\a\, {\tr}\,R^2 +  b_i^\a \left(\frac{2}{\lambda_i} {\tr}\, F_i^2\right)  + 2\, c^\a\, F^2 \right) \ .
\label{Y1}
\ee
Here, $F_i$ is the field strength of the $i^{th}$ component of the gauge group, $F$ denotes the $U(1)_R$ field strength, ${\tr}$ is the trace in the fundamental representation, and summation over $i$ is understood\footnote{There exists the identity $({\tr}\, F^2)/\lambda = 
({\rm Tr}\,F^2)/(2h^\vee)$ where  ${\rm Tr}$ is the trace in the adjoint representation and $h^\vee$ is the dual Coexeter number. $\lambda$ is in fact the index of the fundamental representation of group $G$.},  and $\lambda_i$ is normalization factor which is fixed by demanding that 
the smallest topological charge of an embedded $SU(2)$ instanton is 1 \cite{Bernard:1977nr}. These factors, which  are equal to the Dynkin indices of the fundamental representations, are listed below for the groups needed here. 
\begin{align} 
&G && E_7 && E_6 && F_4 && G_2 && Sp(9)
\nn\\[-6pt]
& \lambda && 12 && 6 && 6 && 2 &&~1
\end{align}

The vectors $(a, b_i,c)$ in the space $\mathbb{R}^{1,1}$ should belong to an integral lattice, referred to as the {\it anomaly lattice}. For the three models we are considering, these vectors are  \cite{RandjbarDaemi:1985wc,Avramis:2005qt,Avramis:2005hc}
\begin{align}
& (A) && a = ( 2,-2)\ , && b_6=(1,3) \ , && b_7 =(3,-9)\ ,  && c=(2,18)\ , 
\label{abc1}\\
& (B) && a = ( 2,2)\ , && b_2=(3,15) \ , && b_7 =(3,1)\ ,  && c=\left(2,-\frac{38}{3}\right)\ ,
\label{abc2}\\
& (C) && a = ( 2,-2)\ , && b_4=(2,-10) \ , && b_9 = \left(1,\frac12 \right)\ ,  && c=(2,19)\ .
\label{abc3}
\end{align}
Note that in models $(A)$ and $(C)$, the anomaly polynomial starts with $-{\tr}\,(R^2)^2$, while in the model $(B)$ it starts with $+{\tr}\,(R^2)^2$. 
The anomaly polynomials for models (A) \cite{RandjbarDaemi:1985wc,RandjbarDaemi:2004qr}, (B) \cite{Avramis:2005qt} and (C) \cite{Avramis:2005hc} are provided in the Appendix.

For groups with non-trivial sixth homotopy group, there may be global anomalies associated with large gauge transformations not connected to the identity. Among the gauge groups of the three models above, only $G_2$ has a nontrivial such homotopy group given by $\pi_6(G_2)= \mathbb{Z}_3$. In that case the vanishing of the global anomaly requires that \cite{Bershadsky:1997sb}
\be
G_2\ : \qquad  1- 4 \sum_{R} n_R\,{d_R} =0\ {\rm mod}\ 3\ ,
\ee
where $n_R$ is the number of (half)hypermultiplets transforming in the representation $R$ of $G_2$, and $d_R$ is defined by ${{\tr}_R F^4} = d_R ({\tr}\,F^2)^2$. Since $n_{14}= \frac12 \times 56$ and $d_{14}=\frac52$ for model (B), the global anomaly is absent \cite{Avramis:2005qt}.

\section{Supersymmetry, bosonic field equations and dyonic strings}

In order to highlight the issues that arise in the context of finding dyonic string solutions of the $U(1)_R$ gauged theory, here we shall review the bosonic field equations, with the assumptions that the hyperscalar fields are set to zero in these equations. We start by introducing a metric $G_{\a\b}$, and $SO(1,1)$ invariant tensor $\Omega_{\a\b}$ in terms of $\Omega$-orthogonal vectors $e_a$ and $j_\a$ as follows
\footnote{We use the notation and conventions of \cite{Monnier:2017oqd} to large extent. One particular exception is that we take $e\cdot H$ to belong to the supergravity multiplet, rather than the tensor multiplet. This is in accordance with the conventions of \cite{Nishino:1997ff},  where the similarity transformation $ S \eta S^T =\Omega$ and $SC_i = {\sqrt 2}\,v_i$ are to be made with $S=\frac{1}{\sqrt 2} \begin{pmatrix} 1 & 1\\ -1 & 1 \end{pmatrix}$ to get to the $SO(1,1)$ basis used here. We have also let $B_{NS} \to {\sqrt 2}\, B_{\rm here}$. }
\bea
G_{\a\b} &=& e_\a e_\b + j_\a j_\b\ ,\qquad \Omega_{\a\b} = -e_\a e_\b + j_\a j_\b \ ,
\nn\w2
e_\a &=&  \frac{1}{\sqrt 2} \big( e^{-\vp},\, -e^{\vp} \big)\ ,\qquad 
j_\a= \frac{1}{\sqrt 2} \big( e^{-\vp},\, e^{\vp} \big)\ .
\eea
They satisfy $e\cdot e=-1\ , j\cdot j = 1$ and $e\cdot j=0$ where the inner product is with respect to $\Omega^{\a\b}= (\Omega_{\a\b})^{-1}$.  It is also convenient to introduce the notation
\be
v_L^\a := \frac12 a^\a\ ,\qquad v_i^\a := \left( \frac{2 b_1^\a}{\lambda_1},\  \frac{2 b_2^\a}{\lambda_2}, \ 2c^\a \right)\ ,
\ee
where $i$ labels the group $G_1 \times G_2 \times U(1)_R$. Then, \eq{Y} can be written as
\be
Y^\a = \frac{1}{16\pi^2} \Big( v_L^\a\, {\tr\,R^2} + v_i^\a\,{\tr}\, F_i^2  \Big) \ ,
\ee
where $ v_3^\a {\tr}\,F^2 = v^\a F^2$. The constant vectors $v_i^\a$ can be directly read of from \eq{abc1}, \eq{abc2} and \eq{abc3}. 
Since $dY^\a=0$, one can locally defined the associated Chern-Simons form through $Y^\a = d\Gamma^\a$. The 3-form field strength is then defined as
\be
H^\a = dB^\a + \alpha'\, \G^\a\ ,\ \ \  {\rm with}\ \ \ d\Gamma^\a = Y^\a\ ,
\label{gm}
\ee
where $\alpha'$ is the `inverse string tension'. 

If we set $v_L^\a=0$, and either $v_i^1=0$ or $v_i^2=0$, then a classically locally supersymmetric and gauge invariant action exists for arbitrary $v_i^1$ or $v_i^2$ \cite{Nishino:1986dc}. If we switch on both $v_i^\a$ simultaneously, local supersymmetric field equations of motion have also been constructed, but  anomalies in gauge transformation and local supersymmetry arise \cite{Nishino:1997ff,Riccioni:1998th}. This is to be expected, since Green-Schwarz counterterms required for the cancellation of one-loop anomalies are present, and therefore the classical and one-loop quantum effects are mixed. The GS counterterm also requires that the parameters $v_L^1$ and $v_L^2$ are turned on, and fixes $v_i^1,$ and $v_i^2$ in terms of a single dimensionful parameter $\alpha'$. Furthermore, supersymmetry is now broken already at order $\a'$, since $R$ and $F$ have the same dimension, and arise in the field equations already at order $\a'$. 

For $v_L^\a=0$, the bosonic field equations with hypermultiplet scalars set to zero, and in Einstein frame, take the form \cite{Nishino:1997ff,Sagnotti:1992qw,Riccioni:2001bg}
\bea
 \star \left( G_{\a\b} H^\b\right) &=& \Omega_{\a\b}\,H^\b\ ,
\label{fe1}\w2
\a'\,D\left(\star\, e\cdot v_i\,F_i\right) &=& - \frac{1}{\sqrt 2} \a'\,v_i^\a\, G_{\a\b} \,\star H^\b \wedge F_i\ ,
\label{fe2}\w2
R_{\mu\nu} &=&  \partial_\mu \vp \partial_\nu \vp + \frac12 G_{\a\b} (H^\a \cdot H^\b)_{\mu\nu}  +2{\sqrt 2}\, \a'\, e\cdot {v}_i\, {\tr}\, \big( (F_i^2)_{\mu\nu} -\frac18 g_{\mu\nu} F_i^2\big)
\nn\\
&&  + \frac{1}{8\sqrt 2}\,\frac{1}{\a'}\, \left(e\cdot v\right)^{-1} g_{\mu\nu}\ ,
\label{fe3}\w2
\nabla_\mu \partial^\mu \vp &=& -\frac{1}{\sqrt 2}\,\a'\,  j\cdot v_i\,{\tr}\,F_i^2  - \frac13 e_\a\, j_\b\, H^\a \cdot H^\b 
 -\frac{1}{4\sqrt 2}\,\frac{1}{\a'}\,\frac{j\cdot v}{(e\cdot v)^2}\ ,
 \label{fe4}
\eea
 with self explanatory meaning of the notations $H^\a\cdot H^\b, (F^2)_{\mu\nu}$ and $F^2$. It follows from \eq{fe1} and \eq{gm} that
\be
d \star \left(G_{\a\b} H^\b \right) = \a'\,\Omega_{\a\b}\, Y^\b\ ,\qquad dH^\a = \alpha'\,Y^\a\ .
\label{fe5}
\ee
Thus $\Omega_{\a\b} Y^\b$ and $Y^\a$ are the electric and magnetic sources, respectively. Note also that $\star e\cdot H=-e \cdot H$ belongs to the supergravity multiplet, and $\star j\cdot H = j\cdot H$ is in the single tensor multiplet.  
We also see from \eq{fe2} that there are terms proportional to $\a'^2$ that break the gauge invariance. Therefore, these equations should be treated as order $\a'$ equations, and thus letting $H \to dB$ in \eq{fe2}, and   $ H\cdot H \to  dB \cdot \left(dB + 2\a' \Gamma\right) $ in \eq{fe3} and \eq{fe4}. 

Turning on $v_L^\a$ breaks supersymmetry even at order $\a'$. This phenomenon has been well studied in particular in $10D$  \cite{Bergshoeff:1989de} and it is known that restoring supersymmetry at order $\a'$ requires the addition of a Riemann curvature-squared term into the action roughly by letting, schematically,  $\a' F^2 \to \a' (R^2 + F^2)$. In the ungauged $6D$ theory, similar phenomenon occurs, and such terms have been considered in \cite{Duff:1996rs} in the context of heterotic-heterotic string duality, and in \cite{Fontanella:2019avn}, in the context of constructing Killing spinors.

Considering the gauged supergravities, while a Noether procedure has not been carried out completely as yet for the full system at order $\a'$, taking into account \cite{Han:1985pp}, we expect the following result in the absence of hypermultiplet 
\bea
S &&= \int \Big\{ \frac14 R(\o) \star {\oneone}  - \frac14 \star d\vp \wedge d\vp -  G_{\a\b} \star (dB^\a) \wedge ( dB^\b + 2 \a'\,\Gamma^\b )  + \frac12 \a'\,\Omega_{\a\b} B^\a \wedge Y^\b
 \nn\\
&& -\frac{1}{4 \sqrt 2}\, \a'\, e\cdot \big( v_L\, {\tr}\, \star R(\o)\wedge R(\o)  +  v_i\, {\tr}\, \star F_i \wedge F_i\big)    
-\frac{1}{8{\sqrt 2}\a'}\,\left(e\cdot v \right)^{-1} \star {\oneone} + \cdots \Big\},
\label{a1}
\eea
where the ellipses are for yet to be determined $H=dB$ and dilaton dependent terms\footnote{In the absence of $U(1)_R$ gauging and if $v_\a^2=0$, then such terms would be accounted for by shifting the spin connection occurring in the ${\rm Riem}^2$ term by torsion as $ \o \to \o-\frac12 H$.}. A similar action for the ungauged theory in string frame, albeit in a non manifestly $SO(1,1)$ invariant form, was given in \cite{Duff:1996rs}. 
In obtaining the field equations from this action, the duality equation  \eq{fe1} is to be imposed {\it after} the variation of the action. With this in mind, it can be checked that this action gives the equations of motion \eq{fe2}-\eq{fe5}, if $v_L^\a$ is set to zero. 
The inclusion of $v_L^\a$ effects will clearly introduce higher derivative terms in the Einstein's and dilaton field equations, though the consequences for the other field equations remain to be investigated, since the Noether procedure for the full system at order $\alpha'$ has not been established as yet. 
As for the term $\int \Omega_{\a\b} B^\a \wedge  Y^\b$ in the above action, naturally it plays a crucial role in the discussion of Dirac quantization of dyonic string charges, as we shall see later. 

Turning to the action \eq{a1}, the requirement that the gauge kinetic terms are ghost-free imposes the constraints $e\cdot v_i >0$ \cite{Sagnotti:1992qw,Seiberg:1996vs,Duff:1996cf}. These kinetic terms for models A, B and C are given by
\be
 A: \quad  -\frac{\a'}{24} \left( e^{-\vp} -3 e^{\vp} \right) {\rm tr}\, \star F_6 \wedge F_6  -\frac{\a'}{16} \left( e^{-\vp} + 3e^{\vp} \right) {\rm tr}\, \star F_7 \wedge F_7 -\frac{\a'}{2} \left( e^{-\vp} -9 e^{\vp} \right) \star F_1 \wedge F_1\ ,  
\nn
\ee
\be
B: \quad -\frac{3\a'}{8} \left(  e^{-\vp} -5 e^{\vp} \right) {\rm tr}\, \star F_2 \wedge F_2  -\frac{\a'}{48} \left( 3e^{-\vp} - e^{\vp} \right) {\rm tr}\, \star F_7 \wedge F_7 -\frac{\a'}{6} \left( 3e^{-\vp} +19 e^{\vp} \right) \star F_1 \wedge F_1\ , 
\nn
\ee
\be
C: \quad -\frac{\a'}{12} \left(  e^{-\vp} +5  e^{\vp} \right) {\rm tr}\, \star F_4 \wedge F_4  -\frac{\a'}{8} \left( 2e^{-\vp} - e^{\vp} \right) {\rm tr}\, \star F_9 \wedge F_9 -\frac{\a'}{4} \left( 2e^{-\vp} -19 e^{\vp} \right) \star F_1 \wedge F_1\ .
\ee
It is easy to check that the positivity condition for these kinetic terms are satisfied for 
\be
(A):\quad e^{-\vp}>3\ , \qquad\qquad (B):\quad  e^{-2\vp} >5\ ,\qquad\qquad
(C):\quad e^{-2\vp}>\ft{19}2\ .
\ee
The perturbative results are reliable for sufficiently negative values of $\vp$, while the lower bounds on $e^{-\vp}$ stated above correspond to the strong Yang-Mills coupling regime. It is also clear that there are a number of values for $\vp$ where some of the Yang-Mills couplings vanish. As discussed in detail in \cite{Duff:1996rs}, these are points where phase transitions are expected to occur.

The last terms in Einstein and dilaton field equations above involve a potential function, and arise as a consequence of the $U(1)_R$ gauging, and that they are absent in the ungauged 6D models, even if the gauge groups include ``external" $U(1)$ factors. These terms clearly have significant impact on the structure of the vacuum as well as the nonperturbative exact solutions. For example, it is easy to check that these terms forbid Minkowski$_6$ and $(A)dS$ vacuum solutions. 

As for dyonic string solutions of $U(1)_R$ gauge theory, to our best knowledge, few solutions exist to equations in which only the classically exactly supersymmetric supergravity equations  are solved. The action with $v_L^\a=0, v_i^2=0$ \cite{Nishino:1986dc} has been used to obtain the dyonic solutions mentioned above. Here $B=B^1$ which represents the combination of the 2-forms residing in supergravity and single tensor multiplet, and therefore it is free from (anti)self duality condition. The action can be read of from \eq{a1} by taking $v_L^\a=0, v_i^2=0, B^1 \equiv B$, and takes the form\footnote{We have let $\vp \to -\vp/{\sqrt 2}$ in the results of \cite{Nishino:1986dc}.}
\bea
S &=& \int \Big( \frac14 R \star {\oneone}  -\frac14 \star d\vp \wedge d\vp - \frac12\,e^{-2\vp}\,G \wedge G 
\nn\\
&& - e^{-\vp} \left(  {v}_{i}\, {\tr}\, \star F_i \wedge F_i  
 +{v}\,\star F \wedge F\right)   -\frac{1}{4 v} \,e^{\vp} \star {\oneone} \Big) \ ,
\label{a2}
\eea
where $G=dB^1 + \a' \Gamma^1 \big|_{v_L=0}$. 
The solution found for the resulting equations have only the following nonvanishing fields \cite{Gueven:2003uw} takes the form
\bea
ds^2 &=& c^2 dx^\mu dx_\mu + a^2 dr^2 + b^2 \left( \sigma_1^2 + \sigma_2^2 + \frac{4g P}{k}\,\sigma_3^2 \right)\ ,
\nn\w2
G &=& P \sigma_1 \wedge \sigma_2 \wedge \sigma_3 - u(r)\, d^2 x \wedge dr\ ,
\nn\w2
F &=& k\, \sigma_1 \wedge \sigma_2\ , \qquad e^{2\vp} = \left(Q_0+\frac{Q}{r^2}\right)\left(P_0+\frac{P}{r^2}\right)^{-1}\ ,  
\eea
where $a,b,c,u$ are functions of $r$ which can be found in \cite{Gueven:2003uw}, $k,P_0,Q_0,P,Q$ are constants, $g$ is the $U(1)_R$ coupling constant, $\sigma_i$ are left-invariant one-forms on the 3-sphere satisfying $d\sigma_i = -\frac12 \epsilon_{ijk} \sigma_j \wedge \sigma_k$. The solution also requires that\footnote{Note that upon letting $A_\mu \to A_\mu/g$, this condition becomes $4P=k\left(1-k(\a')^{-1}\right)$.} 
\be
4 g P = k(1- 2 k g)\ ,
\label{sc}
\ee
which is a condition not arising in the ungauged $6D$ theory, and it has $1/4$ of the $6D$ supersymmetry. It is asymptotic to a cone over (Minkowski)$_2 \times$ squashed $S^3$, as opposed to the expected maximally symmetric known vacuum solution given by Minkowski$_4 \times S^2$ \cite{Salam:1984cj}, and the dilaton blows up asymptotically \cite{Gueven:2003uw}. The near horizon limit of the gauged dyonic string is given by $AdS_3\times$ squashed $S^3$ with fraction of supersymmetry increased from 1/4 to 1/2.
A dyonic string solution of the $U(1)_R$ gauged theory in which an additional $U(1)$ gauge field residing in $E_7$ is activated was found in \cite{RandjbarDaemi:2004qr}, under the assumptions that are similar to those of \cite{Gueven:2003uw} outlined above. In particular, $1/4$ supersymmetry also arises and again the dilaton blows up asymptotically.   
 
\section{Constraints on anomaly coefficients}

The factorization of the anomaly polynomial has been shown to imply that \cite{Kumar:2010ru}
\be
a\cdot a\ ,\quad a\cdot b_i\ ,\quad b_i \cdot b_j\ \in\ \mathbb{Z}\quad \mbox{for all}\ i, j\ ,
\label{p1}
\ee
where the products are in $\mathbb{R}^{1,1}$ with metric $\Omega_{\a\b}$. 
The condition above can be checked explicitly for all three models studied here. 
The fact that the anomaly coefficients belong to an integral lattice is not sufficient for the consistency of the theory. To elaborate further on this point, it is convenient to first re-express the form $Y^\a$ appearing in the Bianchi identity $dH^\a = \alpha'\,Y^\a$ in terms of characteristic forms, applied to the models considered here taking the form \cite{Monnier:2017oqd,Monnier:2018cfa}
\be
Y^\a = \frac14 a^\a p_1 -b_i^\a\,c_2^i + \frac12\, c^\a\,(c_1)^2\ ,
\label{Y}
\ee
where $p_1, c_2$ and $c_1$ are the Chern-Weyl representatives of the indicated cohomology classes defined as 
\be
p_1 = \frac{1}{8\pi^2} {\tr}\, R^2\ ,\qquad c_2^i = -\frac{1}{8\pi^2} \left( \frac{1}{\lambda_i} {\tr}_i\,F^2 \right)\ ,\qquad c_1 = \frac{F_1}{2\pi}\ .
\ee
It is then argued in \cite{Monnier:2017oqd} that the string charge defined by the integral $\int_{\Sigma_4} Y$, where $\Sigma_4$ is any integral 4-cycle, must be cancelled by background self-dual strings. Consequently, it is argued that this charge must yield an element of the {\it unimodular string charge lattice} $\Lambda_S$, and this ``string quantization condition" is explicitly stated as\footnote{The anomaly coefficients are measured in units of $\alpha'$ which we set equal to one.}
\be
\int_{\Sigma_4} Y \ \in \Lambda_S\ .
\label{QC}
\ee
The fact that $\Lambda_S$ is a unimodular, equivalently self-dual, lattice can be seen from basic arguments that can be found, for example, in \cite{Seiberg:2011dr} and \cite{Monnier:2017oqd}.

The completeness hypothesis was taken a step further by Monnier, Moore and Park \cite{Monnier:2017oqd} who assumed that a consistent supergravity theory may be put on an arbitrary spin manifold and that any smooth gauge field configurations should be allowed in the supergravity ``path integral''. 
The strategy employed in \cite{Monnier:2017oqd} is then to assume the generalized completeness hypothesis and obtain strong constraints by evaluating \eq{QC} on suitable chosen spacetimes $M$ and gauge bundles. In particular, taking $M=CP^3$, and evaluating \eq{QC} along a suitable 4-cycle, the derive the condition (applied to the groups considered here)
\cite{Monnier:2017oqd}
\be
a,\ b_i,\ \frac12 c \ \in \ \Lambda_S\ ,\qquad \Lambda_S\ unimodular\ .
\label{mc1}
\ee
A special case of this condition was derived earlier by Seiberg and Taylor \cite{Seiberg:2011dr} in the form
\be
b_i \ \in \ \Lambda_S\ , \qquad \Lambda_S\ unimodular\ .
\label{ST}
\ee
by demanding consistency of the theory by means of Dirac quantization of charges, once it is compactified on various spaces, such as $T^2, T^4$ and $CP^2$. It was also argued that the presence or absence of the Abelian factors in the gauge group does not effect their results, which depends only on the non-Abelian part of the gauge group.

In \cite{Monnier:2017oqd}, it has also been shown that the constraint \eq{mc1} is  equivalent to the statement \cite{Monnier:2017oqd}
\be
 a \in \Lambda_S\ ,\quad \frac12\mathfrak{b} \in H^4 (B{\widetilde G}; {\mathbb Z}) \times\Lambda_S\ , \qquad \Lambda_S\ unimodular\ .
\label{mc11}
\ee
where $B{\widetilde G}$ is the classifying space of the universal cover of the semisimple part of the gauge group $G$. The bilinear form $\mathfrak{b}$ in our case, where $G=\bigotimes_i G_i\times U(1)_R$ with $i=1,2$, can be written as   
\be
\mathfrak{b} =\bigoplus_i b_iK_i\oplus c\ .
\ee
Here $b_i$ is the anomaly coefficient associated with the non-abelian Chern form ${\tr}\, F^2_i$ and $K_i$ is the canonically normalized  Killing form\footnote{Here we use standard math convention in which $K_i$ is unit matrix of dimension spanning the rank of the underlying Lie algebra, upon its restriction to the Cartan subalgebra.}, with respect to which, the length squared of the longest simple root is 2 (for U(1), the root length squared is 1). 
It has also been shown that this is equivalent to the statement that $\mathfrak{b} $ is an even $\Lambda_S$-valued bilinear form when restricted to the {\it coroot lattice} \cite{Monnier:2017oqd}. Specifically, \eqref{mc11} implies for any $x\,,y$ inside the coroot lattice, 
\be
\ft12\mathfrak{b}(x,x)\in \Lambda_S\,,\quad {\rm and} \quad \mathfrak{b}(x,y)\in \Lambda_S\quad {\rm for}
\quad x\neq y\,.
\label{mc12}
\ee
Taking into account the the global structure of the gauge group, the condition \eq{mc11} has been strengthen to \cite{Monnier:2017oqd}
\be
a \in \Lambda_S\ ,\quad \frac12 \mathfrak{b} \in H^4 (BG; {\mathbb Z}) \times \Lambda_S\ , \qquad \Lambda_S\ unimodular\ .
\label{mc2}
\ee
which leads to conditions similar to \eqref{mc12} with $x\,,y$ now belonging to the cocharacter lattice. For a detailed description of various lattices of $G$, see \cite{Gukov:2006jk}. We only emphasize the following key aspects here. 
There is a general relation among the coroot lattice, cocharacter lattice and the coweight lattice for a given semisimple Lie algebra $\mathfrak{g}$ \cite{Gukov:2006jk,Simon} 
\be
\L^{\rm coroot} \subseteq \L^{\rm cocharacter} \subseteq \L^{\rm coweight}\ . 
\label{rrr}
\ee
These inclusions are determined by the global structure of the group $G$. Specifically, \cite{Gukov:2006jk,Simon}
\be 
\L^{\rm cocharacter}/\L^{\rm coroot}=\pi_1(G)\,,\quad \L^{\rm coweight}/\L^{\rm cocharacter}=Z(G)\ ,
\label{hr}
\ee
where $\pi_1(G)$ is the first homotopy group of $G$ and $Z(G)$ denotes the center of G. 
For connected Lie groups, $H^4 (BG; {\mathbb Z})$ is torsion free. For disconnected groups, there could potentially be a torsion class whose coefficient should be quantized in terms of the string charge lattice \cite{Monnier:2018nfs}.

As mentioned in the introduction, Monnier and Moore extended the above considerations and arrived at a stronger criterion by seeking the conditions under which the GS counterterm is well defined. This leads to the requirement for the existence of a topologically trivial field theory in 7D, referred to as Wu-Chern-Simons theory, and a set of conditions for the 6D theory to be free from all anomalies. Applied to the case under consideration, where the gauge groups are connected, the proposition states that given string charge lattice $\Lambda_S$, and the anomaly polynomial $A_8$, and 4-form $Y$ as defined in\eq{Y}, assume that \cite{Monnier:2018nfs}

\begin{enumerate}
\item $A_8=\frac12 Y\wedge Y$;
\hfill\refstepcounter{equation}\textup{(\theequation)}\label{pp1}
\item $\Lambda_S$  is unimodular;
\hfill\refstepcounter{equation}\textup{(\theequation)}\label{p2}
\item ${\mathfrak b} \in 2H^4(BG;\Lambda_S)$;
\hfill\refstepcounter{equation}\textup{(\theequation)}\label{p3}
\item $a\in \Lambda_S$ is a characteristic element;
\hfill\refstepcounter{equation}\textup{(\theequation)}\label{p4}
\item $\Omega_7^{\rm Spin} (BG)=0$\ , 
\hfill\refstepcounter{equation}\textup{(\theequation)}\label{p5}
\end{enumerate}
where $\Omega_7^{\rm Spin} (BG)=0$ is the spin cobordism group associated with Lie group G. 
Then all anomalies of the 6D theory, local and global, cancel. The ways in which this proposition extends \eq{mc12} are as follows. Firstly, the derivation of the third condition does not rely on the generalized completeness hypothesis. Furthermore, the fourth condition states not only that $a\in \Lambda_S$ but it is also a characteristic element. Finally, the fifth condition clearly goes beyond what is required in \eq{mc12}.

\section{Application of the consistency conditions}

The Monnier and Moore also tacitly assumes that string defects are included whenever they are necessary to satisfy the tadpole condition, and that their worldsheet anomalies cancel the boundary contributions to the anomaly of the supergravity theory through the anomaly inflow mechanism, as has been stated in \cite{Monnier:2018nfs}. Very recently, {\cite{Kim:2019vuc} proposed that using the gravitational and gauge anomaly inflow on the probe string, one can compute the worldsheet gravitational central charge and the gauge group's current algebra level depending on the string charge and the bulk anomaly coefficients. (For earlier work in this context, see \cite{Kim:2016foj,Shimizu:2016lbw}.) The requirement that the left-moving central charge should be large enough to allow the unitary representations of the current algebra for a given level imposes a constraint on the allowed gauge group content. However, the fact that the near horizon limit of the gauged dyonic string is given by 1/2 BPS $AdS_3\times$ squashed $S^3$ suggests that the IR CFT of the probe string coupled to the gauged supergravity should be a two-dimensional ${\cal N}=2$ CFT, in contrast to \cite{Kim:2019vuc} where the  worldsheet IR CFT is described by a (0,4) CFT. Thus one cannot directly apply the result of \cite{Kim:2019vuc} here before a careful study on the low energy dynamics of the probe string is carried out. Altogether, whether the tacit assumptions made as prelude to the Monnier-Moore proposition are satisfied by the $U(1)_R$ gauged $6D (1,0)$ supergravities is not entirely clear,and remain to be investigated. Nonetheless, we shall at least assume that suitably behaved dyonic string solutions exist and proceed below with the analysis of the consequences of the above proposition for these models. 

To begin with, condition 1 is obviously satisfied by Models A,B, and C. Next, we look at condition 5. To this end, we note that\footnote{We are very grateful to I. García-Etxebarria and M. Montero, for explaining to us their results for $\Omega_7^{\rm Spin} (BG)$ for $G_2$ (unpublished), $F_4, E_6$ and $E_7$, and $E_6/{\mathbb Z}_3$ (unpublished).} 
\bea
&& \Omega_7^{\rm Spin} (BG_2)=0\ ,\qquad \Omega_7^{\rm Spin} (BF_4)=0\ ,\qquad \Omega_7^{\rm Spin} (BE_7)=0\ ,\qquad \Omega_7^{\rm Spin} (BSp(9))=0\ ,
\nn\w2
&& \Omega_7^{\rm Spin} (BE_6/{\mathbb Z}_3)= D_3\ ,
\label{br}
\eea
where $D_3$ is yet to be determined group of exponent $6$. Since it is not known yet whether $D_3$ is trivial or not, we shall examine the other conditions required by the proposition in the case of Model A which has the symmetry $(E_6/{\mathbb Z}_3)\times E_7\times U(1)_R$. As for models B and C, given the results \eq{br}, they pass the 5th condition of the proposition. 

For the convenience of further discussion, we introduce the notation
\be
{\cal M}(x,y)=\begin{pmatrix} x\cdot x & x\cdot y  \\ y\cdot x & y\cdot y  \end{pmatrix}\ , 
\ee
where $x\,,y$ are $\mathbb{R}^{1,1}$ vectors and the product is defined with respect to $\Omega_{\a\b}$. The fact that string charge lattice $\L_S$ is unimodular implies that $-{\rm det}{\cal M}(x,y)$ must be a square of a positive integer for any $x\,,y\in \L_S$. 

In using the relations \eq{rrr}, it is also useful to note that as far as the nonabelian groups appearing in models A, B and C are concerned, $\pi_1(G)=\oneone$ and $Z(G)=\oneone$ for all, except that 
\be
\pi_1(E_6/\mathbb{Z}_3) = \mathbb{Z}_3\ ,\qquad Z(E_6)=\mathbb{Z}_3\ ,\qquad Z(E_7)=\mathbb{Z}_2\ , \qquad Z (Sp(9))=\mathbb{Z}_2\ .
\ee
In the following, we will test constraints stated in the proposition for models A, B and C, though, we will also see if only the weaker constraints \eqref{mc1} and/or \eqref{mc2} are satisfied in some cases.

\subsection{The $ \left(E_6/{\mathbb{Z}_3}\right) \times E_7 \times U(1)_R$ invariant model}

We first compute $-{\rm det}{\cal M}(x,y)$ for $x\,,y$ being any two distinct $\mathbb{R}^{1,1}$ vectors among $a\,,b_6\,,b_7\,,\ft12c$. The result is given by
\bea
-{\rm det}{\cal M}(a,b_6)&=&8^2\,,\quad -{\rm det}{\cal M}(a,b_7)=12^2\,,\quad
-{\rm det}{\cal M}(a,\ft12c)=20^2\ , \nn\w2
-{\rm det}{\cal M}(b_6,b_7)&=&18^2\,,\quad -{\rm det}{\cal M}(b_6,\ft12c)=6^2\,,\quad -{\rm det}{\cal M}(b_7,\ft12c)=36^2\ .
\eea
 therefore the anomaly coefficients in this model are compatible with the second condition of the proposition, namely  with \eq{p2}, that the anomaly coefficients lie on a unimodular string charge lattice. To verify that the lattice is indeed unimodular, we proceed by choosing as a basis of a unimodular charge lattice
\be
e_1=(1,0)\,,\quad e_2=(0,1)\ ,
\ee
and observe that the anomaly coefficients can be recast as linear combination of $e_1\,,e_2$ with integer coefficients. Note that this lattice is {\it even}.

Next, we inspect the anomaly coefficients against the stronger constraint \eqref{mc2}. In order to do so, we need to evaluate the bilinear form $\mathfrak{b}$ on the cocharacter lattice of $E_6 \times E_7 \times U(1)_R$. 
In the model with gauged $\left(E_6/{\mathbb{Z}_3}\right) \times E_7 \times U(1)_R$ symmetry, $E_6$ appears only in the adjoint representation. Therefore, a vector $v$ on the $E_6$ cocharacter lattice should satisfy
\be
e^{2\pi{\rm i} v_ih_i }=\oneone_{78\times 78}\ ,
\ee
where $h_i\,,i=1,\cdots 6$ are generators of Cartan subalgebra (in the Cartan-Weyl basis) of $E_6$ in the adjoint representation. Clearly, such $v$ lies in the coweight lattice of $E_6$ spanned by $\check{w}_m$ that obey $\sum_i(\check{w}_m)_i(r_n)_i=\d_{mn}$, for simple roots labeled by $r_n$. Using the definition of coweights $\check{w}_m$, we can evaluate the bilinear form $K_6$ on the coweight lattice and obtain
\be
K_6(\check{w}_r,\check{w}_s)=\frac13\left(
\begin{array}{cccccc}
 4 & 5 & 6 & 4 & 2 & 3 \\
 5 & 10 & 12 & 8 & 4 & 6 \\
 6 & 12 & 18 & 12 & 6 & 9 \\
 4 & 8 & 12 & 10 & 5 & 6 \\
 2 & 4 & 6 & 5 & 4 & 3 \\
 3 & 6 & 9 & 6 & 3 & 6 \\
\end{array}
\right)\,,
\ee
which is equal to the inverse of the $E_6$ Cartan matrix. This happens to be so because Lie algebra of $E_6$ is simply laced and thus the length squared of every simple root equals 2, implying the coweight vector coincides with the fundamental weight vector. From the expression above, we single out a particular element $K_6(\check{w}_1\,,\check{w}_1)$ whose product with $b_6$ leads to the following vector on $\mathbb{R}^{1,1}$ %
\be
\tilde{b}_6=\ft12b_6K_6(\check{w}_1\,,\check{w}_1)=\ft23b_6\ .
\ee
This gives $-{\rm det}{\cal M}(a,\tilde{b}_6)=(\ft{16}3)^2$, which means that $\tilde{b}_6$ and $a$ cannot belong to the same unimodular lattice. Thus, the third condition of the proposition, namely \eq{p3}, is not satisfied.

\subsection{The $E_7 \times G_2 \times U(1)_R$ invariant model}

Similar to the previous case, we first investigate whether the anomaly coefficients can be embedded in a unimodular lattice, by computing $-{\rm det}{\cal M}(x,y)$ for $x\ ,y$ being any two distinct $\mathbb{R}^{1,1}$ vectors among $a\,,b_2\,,b_7\,,\ft12c$. It turns out that
\bea
-{\rm det}{\cal M}(a,b_2)&=&24^2\,,\quad -{\rm det}{\cal M}(a,b_7)=4^2\,,\quad
-{\rm det}{\cal M}(a,\ft12c)=(\ft{44}3)^2\ , \nn\w2
-{\rm det}{\cal M}(b_2,b_7)&=&42^2\,,\quad -{\rm det}{\cal M}(b_2,\ft12c)=34^2\,,\quad -{\rm det}{\cal M}(b_7,\ft12c)=20^2\ .
\eea
As $-{\rm det}{\cal M}(a,\ft12c)$ is not given by a positive integer squared, the anomaly coefficients $a\,,b_2\,,b_7\,,\ft12c$ cannot all belong to a unimodular lattice, thus violating the second condition of the proposition, namely \eq{p2}. 

\subsection{The $F_4 \times Sp(9) \times U(1)_R$ invariant model}

For this model, we obtain $-{\rm det}{\cal M}(x,y)$ for $x\,,y$ being any two distinct $\mathbb{R}^{1,1}$ vectors among $a\,,b_4\,,b_9\,,\ft12c$ as  
\bea
-{\rm det}{\cal M}(a,b_4)&=&16^2\,,\quad -{\rm det}{\cal M}(a,b_9)=3^2\,,\quad
-{\rm det}{\cal M}(a,\ft12c)=21^2\ , \nn\w2
-{\rm det}{\cal M}(b_4,b_9)&=&11^2\,,\quad -{\rm det}{\cal M}(b_4,\ft12c)=29^2\,,\quad -{\rm det}{\cal M}(b_9,\ft12c)=9^2\ ,
\eea
which shows that the necessary condition for the the lattice $\Lambda_S$ being unimodular is satisfied. To establish that it is indeed unimodular, we proceed as follows. We choose the following basis for a unimodular charge lattice 
\be
e_1=(2,0)\,,\quad e_2=(1,\ft12)\ ,
\label{m3basis}
\ee
Next, we observe that the anomaly coefficients in 
this model can be expressed as as linear combinations of $e_1\,,e_2$ with integer coefficients. This shows that condition \eqref{mc1} is indeed satisfied. Note also that the lattice here is {\it odd}, since $e_2\cdot e_2=1$.  
Furthermore, in this model, the group $F_4$ appears only in the adjoint representation,
whereas the hypermultiplet carries also the fundamental representation of $Sp(9)$.
One should also note that since the hyperfermions are singlet under $U(1)_R$, 
it is not possible to form an identity by combining a center element of $Sp(9)$ with an element of $U(1)_R$\footnote{This is different from the $U(2)$ example studied in \cite{Monnier:2017oqd}, where element of the cocharacter lattice is formed by combining a center element of $SU(2)$ with an element of the remaining $U(1)$.}. Since $Z(F_4)$ and $\pi_1(F_4)$ are all trivial, the coroot, cocharacter and coweight lattices are equivalent \eqref{hr}, the third condition of the proposition \eq{p3} reduces to the condition \eqref{mc1}, which we have shown above to be satisfied. 

We now move on to discuss the stronger constraint imposed on $Sp(9)$. 
We recall that $Z(Sp(9))=\mathbb{Z}_2$ and $\pi_1(Sp(9))=\oneone$. Thus the cocharacter lattice
is different from coweight lattice but coincides with the coroot lattice. Indeed the transformation matrix from the standard coroot basis to the standard cocharacter basis is given by the unimodular matrix
\bea
{\cal T}_9=\small{\left(
\begin{array}{ccccccccc}
 1 & 1 & 1 & 1 & 1 & 1 & 1 & 1 & 1 \\
 1 & 2 & 2 & 2 & 2 & 2 & 2 & 2 & 2 \\
 1 & 2 & 3 & 3 & 3 & 3 & 3 & 3 & 3 \\
 1 & 2 & 3 & 4 & 4 & 4 & 4 & 4 & 4 \\
 1 & 2 & 3 & 4 & 5 & 5 & 5 & 5 & 5 \\
 1 & 2 & 3 & 4 & 5 & 6 & 6 & 6 & 6 \\
 1 & 2 & 3 & 4 & 5 & 6 & 7 & 7 & 7 \\
 1 & 2 & 3 & 4 & 5 & 6 & 7 & 8 & 8 \\
 1 & 2 & 3 & 4 & 5 & 6 & 7 & 8 & 9 \\
\end{array}
\right)}\ .
\eea
Thus again, the third condition of the proposition \eq{p3} becomes equivalent to the condition \eqref{mc1} already shown to be satisfied by the explicit construction of the string charge lattice basis given in \eqref{m3basis}. Using \eqref{m3basis} the basis, an element on the charge lattice can be parameterised as 
\be
x=(2n+m\,,\ft12 m)\,,\quad m\,,n\in\mathbb{Z}\,.
\ee
Thus one can easily show that 
\be
a\cdot x=x\cdot x\quad {\rm mod}\quad 2\,.
\ee
Given also that  $\Omega_7^{\rm spin}(BG)=0$ for $G=F_4 \times Sp(9)\times U(1)_R$, we see that all the conditions of the proposition, namely \eq{pp1}-\eq{p5} are satisfied, and therefore this model is free from all anomalies.

\section{Conclusions}

We have highlighted the significance of R-symmetry gauging in $6D, {\cal N}=(1,0)$ supergravity, and focused on three such models that stand out in their accommodation of Green-Schwarz mechanism for the cancellation of all local anomalies in a nontrivial way. We have examined constraints imposed on the anomaly coefficients that are associated with the factorized  anomaly polynomials in these models, as proposed in their strongest form by Monnier and Moore \cite{Monnier:2018nfs}. 
Adopting the assumptions made by these authors, we have found that only Model C, based on the gauge group $F_4 \times Sp(9)\times U(1)_R$, satisfies all the conditions required for freedom from all anomalies, local and global. We have also seen that Model A based on the gauge group $\left(E_6/{\mathbb Z}_3\right)\times E_7 \times U(1)_R$ does have a unimodular lattice, thus satisfying the weaker version of the consistency conditions on the anomaly coefficients \cite{Seiberg:2011dr}, it fails the stronger conditions of \cite{Monnier:2018nfs}, and \cite{Monnier:2017oqd}.

A word of caution is appropriate in applying the Monnier-Moore criteria to the R-symmetry gauged 6D supergravities for the following reason. It is assumed that dyonic strings with proper behaviour that give well defined string charge lattice exists. On the other hand, the existence of dyonic string excitations in these models are yet to be firmly established. The task is primarily complicated by the fact that  the $U(1)_R$ gauging gives rise to a potential function which  effects in a significant way the solution space, and in particular the asymptotic behaviour. The potential comes with an inverse power of $\alpha'$, and certain dyonic string solutions in presence of a potential, and in which the anomaly coefficients $v_\a^2$ (arising in the source term in the $2$-form field equation) are set to zero, \cite{Gueven:2003uw,RandjbarDaemi:2004qr} require a relation among the parameters not seen in the usual dyonic string solutions of the ungauged $6D$ supergravities. Search and in depth study of the dyonic strings solutions of R-symmetry gauged $6D\, (1,0)$ supergravities is needed before a robust conclusion can be reached with regard to their global anomalies. In particular, the consequences for the existence of a worldsheet theory, and the attendant inflaw anomalies require scrutiny, as they may impose yet further constraints on the consistency of the anomaly coefficients, as has been found to be the case for certain {\it ungauged} $6D$ supergravities with minimal supersymmetry \cite{Kim:2019vuc}. 

Notwithstanding the caveat mentioned, we conclude by noting that it is still remarkable that the R-symmetry gauged model with $F_4 \times Sp(9)\times U(1)_R$ satisfies all the constraints of the Monnier-Moore proposition, which  are most stringent ones known as yet. As such, it certainly deserves a closer look, to address further questions such as their place in the arena of swampland conjectures, even though, being conjectures, they are not as firm as the requirement of anomaly freedom so far. It would also be interesting to explore the dyonic string solutions and the charges they are allowed to carry, which can serve as a consistency check to the proposed charge lattice \eqref{m3basis} implying that the minimal charge carried by a purely electric string (labelled by $e_1$ \eqref{m3basis}) is twice as many as that of a purely magnetic string (labelled by $2e_2-e_1$ \eqref{m3basis}). 
A study of $\alpha'$ corrections due to supersymmetry, likely combined with other considerations such as unitarity and causality, may shed some light on the UV completion of the theory, if such a completion exists at all. Finally, it would be interesting to explore the application of the model to cosmology, as it may yield significantly different results compared to those of standard string cosmology, in view of the positive potential afforded by  the R-symmetry gauging.

\section*{Acknowledgements} 

We thank Eric Bergshoeff, Stanley Deser, Andr\'e Henriques, Alex Kehagias and Yoshiaki Tanii for useful discussions. We are very grateful to Greg Moore for explaining his results with S. Monnier and D.S. Park on anomaly constraints, and Inaki Garcia-Etxebarria and Miguel Montero for explaining their results on spin bordism groups. We also thank Istanbul Technical University for hospitality during our visit. The work of Y.P. is supported by a Newton International Fellowship NF170385 of Royal Society, and the work of E.S. is supported in part by  NSF grant PHY-1803875.

\begin{appendix}

\section{The anomaly polynomials}

The fields that contribute to gravitational, gauge and mixed anomalies in $n_T=1, {\cal N}=(1,0)$ supergravity with gauge group $G=G_1\times G_2\times U(1)_R$ in $6D$ are as follows:
\be
\psi_{\mu^A+}\ ,\quad \chi_{-}^A\ ,\quad \lambda^{IA}_{+}\ ,\quad \psi^{aa'}_{-}
\ee
with chiralities denoted by $\pm$. The fermions are symplectic Majorana Weyl, the index $A=1,2$ labels the $SO(2)_R\subset Sp(1)_R$ fundamental, $I$ labels the adjoint representation of the group $G$, and $(aa')$ label the representation content of the hyperfermions under $G_1\times G_2$.

\subsection*{\bf The $\left(E_6/{\mathbb{Z}_3}\right) \times E_7 \times U(1)_R$ model}

From \cite{RandjbarDaemi:1985wc} we have
\bea
Y^1 &=& \frac{1}{16\pi^2}\left(  \tr\, R^2 +\frac13 \tr\, F_6^2 +\frac12 \tr\, F_7^2  + 4 F_1^2 \right)\ ,
\nn\w2
Y^2 &=& -\frac{1}{16\pi^2} \left( \tr\, R^2 -\tr\, F_6^2 +\frac32 \tr\, F_7^2 - 36 F_1^2 \right)\ ,
\eea
The computation of the anomaly polynomial can be found in \cite{RandjbarDaemi:1985wc} where the details of the computation are spelled out\footnote{ See also \cite{RandjbarDaemi:2004qr}, where few typos were corrected in the expressions for the individual contributions to the anomaly polynomial, without any effect on the total and, of course, its factorization.} The generators of the gauge group are taken to be hermitian, and the strength of the $U(1)_R$ coupling constant to be unity, i.e. $D_\mu = \partial_\mu -iA_\mu$. It should be noted that the normalizations in $Y^\a$ are taken differently in various papers. However, following \cite{Monnier:2017oqd}, we take them to be $1/(16\pi^2)$, motivated by the fact that this is the appropriate normalization in the integrals $\int_{\Sigma_4}Y$ discussed in Section 4, in which these integrals are related to Chern-Weyl classes. The freedom to do so stems from the fact that the anomaly coefficients are fixed in terms of $\alpha'$ which we can normalize appropriately, and set equal to one, after having done so. 

\subsection*{\bf The $G_2 \times E_7 \times U(1)_R$ model}

From \cite{Avramis:2005qt} we have
\bea
Y^1 &=&\frac{1}{16\pi^2}\left( \tr\, R^2 +3\tr\,F_2^2 +\frac12 \tr\,F_7^2+4 F_1^2\right)\ ,
\nn\\
Y^2 &=& \frac{1}{16\pi^2}\left(\tr\,R^2 +15\tr\,F_2^2 +\frac16 \tr\,F_7^2 -\frac{76}{3} F_1^2\right)\ .
\eea
The details of the computations for this anomaly polynomial are provided  in \cite{Avramis:2005qt}, where the generators of the gauge group are taken to be ant-hermitian, while here we are employing hermitian generators. The $U(1)_R$ covariant derivative $D_\mu = \partial_\mu -iA_\mu$ is assumed.

\subsection*{\bf The $F_4 \times Sp(9) \times U(1)_R$ model}

From the data provided in \cite{Avramis:2005hc} we find
\bea
Y^1 &=& \frac{1}{16\pi^2}\left(\tr\,R^2 +\frac23\,\tr\,F_4^2 + 2\,\tr\,F_9^2 +4\,F_1^2\right)\ ,
\nn\w2
Y^2 &=& -\frac{1}{16\pi^2}\left(\tr\,R^2 +\frac{10}{3}\,\tr\,F_4^2 - \,\tr\,F_9^2 -38\,F_1^2\right)\ .
\label{m3f}
\eea
As hyperinos transform as ${\bf (52,18)_0}$ under $F_4 \times Sp(9)$, and are neutral under $U(1)_R$, the contributions to the gravitation, gauge and mixed anomalies to the anomaly polynomial are
\bea
P(\psi_\mu ) &=&  \frac{245}{360} \tr\,R^4 -\frac{43}{288} \left(\tr\,R^2\right)^2 -\frac{19}{6} F_1^2\, \tr\,R^2 + \frac{10}{3} F_1^4\ ,
\w2
P(\chi ) &=&  -\left(\,\frac{1}{360} \tr\,R^4 +\frac{1}{288} \left(\tr\,R^2\right)^2\,\right) 
- \frac16 F_1^2\, \tr\,R^2 - \frac{2}{3} F_1^4\ ,
\w2
P(\psi^{aa'}) &=& -\frac12 (52\times 18) \left(\frac{1}{360} \tr\,R^4 +\frac{1}{288} \left(\tr\,R^2\right)^2 \right) 
-\frac12\times \frac16 \left(18\, {\rm Tr}\,F_4^2 +52\, \tr\,F_9^2\right) \tr\,R^2
\nn\\
&& -\frac12 \times \frac23\, \left( 18\,{\rm Tr}\, F_4^4 +52\,\tr\,F_9^4 \right)
-\frac12 \times 4\, {\rm Tr}\,F_4^2 \tr\, F_9^2\ ,
\w2
P(\lambda ) &=& (52+171+1) \left(\,\frac{1}{360} \tr\,R^4 +\frac{1}{288} \left(\tr\,R^2\right)^2 + \frac16 F_1^2\, \tr\,R^2  +\frac23 F_1^4 \,\right)  
\nn\w2
&& +\frac16\,{\rm Tr}\,F_4^2\,\tr\,R^2 +\frac16\,{\rm Tr}\,F_9^2\,\tr\,R^2 
+\frac23\left( {\rm Tr}\,F_4^4 + {\rm Tr}\,F_9^4\right)
\nn\w2
&& +4\, F_1^2\, {\rm Tr}\,F_4^2 +4\, F_1^2\, {\rm Tr}\,F_9^2 
\eea
where ${\rm Tr}$ and $\tr$ denote the traces in the adjoint and fundamental representations, respectively. Here, the group generators are  taken to be hermitian, and for $U(1)$ we have $D_\mu = \partial_\mu -iA_\mu$, and $F_4, F_9, F_1$ are associated with $F_4\times Sp(9)\times U(1)_R$. Using the relations
\bea
{\rm Tr}\,F_4^2 &=& 3\,\tr\,F_4^2\ ,\qquad {\rm Tr}\,F_4^4 = \frac{5}{12}\,\left(\tr\,F_4^2\right)^2\ ,
\w2
{\rm Tr}\,F_9^2 &=& 20\,\tr\,F_9^2\ ,\qquad {\rm Tr}\,F_9^4 = 26\,\tr\,F_9^4 + 3\,\left(\tr\,F_9^2\right)^2\ ,
\eea
the sum $I _8$ becomes
\bea
I_8 &=& - \left(\tr\,R^2\right)^2 +34 F_1^2\,\tr\,R^2 + 152\,F_1^4 - 4\,\tr\,F_4^2\,\tr\,R^2  - \tr\,F_9^2\,\tr\,R^2
\nn\w2
&& -\frac{20}{9}\,\left(\tr\,F_4^2\right)^2  +2\,\left( \tr\,F_9^2\right)^2
-6\,\tr\,F_4^2\, \tr\, F_9^2 +12\,F_1^2\,\tr\,F_4^2 + 80\,F_1^2\,\tr\,F_9^2\ .
\eea
Arranging this data into a $4\times 4$ matrix, it has rank 2, and it factorizes as in \eq{m3f}.

\end{appendix}

\end{document}